\documentclass[
    aip,
    jcp,
    amsmath,
    amssymb,
    reprint, twocolumn, floatfix
]{revtex4-2}

\usepackage{graphicx}
\usepackage{amsmath,units}
\usepackage{mathrsfs}
\usepackage{color} 
\usepackage{xcolor} 
\usepackage[export]{adjustbox}
\usepackage{ulem}
\usepackage{appendix}

\begin{document}
\title{Temporal Correlations and Optimal Feedback Control of an Active Information Engine}

\author{Laura Hoek}
\affiliation{School of Chemistry, Faculty of Exact Sciences, Tel Aviv University, Tel Aviv 6997801, Israel}

\author{Yael Roichman}
\email{roichman@tauex.tau.ac.il}
\affiliation{School of Chemistry, Faculty of Exact Sciences, Tel Aviv University, Tel Aviv 6997801, Israel}
\affiliation{School of Physics \& Astronomy, Faculty of Exact Sciences, Tel Aviv University, Tel Aviv 6997801, Israel}
\affiliation{Center for the Physics and Chemistry of Living Systems. Tel Aviv University, 6997801, Tel Aviv, Israel}

\date{October 2026}

\begin{abstract}
How much does a feedback controller need to know about the fluctuations it exploits? We address this question using a macroscopic information engine driven by self-propelled bristle-robots. Experimentally measured occupancy time series show that temporal correlations strongly affect vacancy detection. At short measurement intervals, mean first-passage times exceed independent-measurement predictions by more than an order of magnitude. A first-order Markov approximation captures much of this effect. It does not reproduce the full first-passage distribution, revealing higher-order temporal memory. Nevertheless, Markov-selected protocols come within 0.15\% of the maximum experimental power. Neglecting correlations causes power losses of several percent and favors larger compression distances. Introducing a sensing cost shifts the optimal strategy toward sparse sensing. The independent model cannot predict this shift, whereas the Markov model captures it. Cross-validation on separate time segments confirms the near-optimal performance of Markov-selected protocols. These findings demonstrate that near-optimal feedback control in correlated nonequilibrium systems can require a simpler statistical description than predicting their full temporal dynamics.
\end{abstract}

\maketitle

\section{Introduction}
Information engines use measurements and feedback to exploit fluctuations and extract work. Originating from the thought experiments of Maxwell's demon~\cite{maxwell2012theory} and Szil\'ard's engine~\cite{szilard1929entropieverminderung,szilard1964decrease}, these machines demonstrate that information acquired through measurement can serve as a thermodynamic resource. Information engines exploit this principle, using feedback loops to rectify fluctuations and extract work, with the amount of extractable work bounded by the information gained~\cite{parrondo_thermodynamics_2015, sagawa2009minimal, sagawa2010generalized}. In recent years, advances in experimental techniques have enabled experimental realizations of information engines in a wide range of platforms~\cite{goerlich2025experimental, admon_experimental_2018, hoek2026piston, Omer_2023, lagoin_human-scale_2022, remi_now, paneru_efficiency_2020, ribezzi-crivellari_large_2019, saha_maximizing_2021, paneru2022colossal, saha_bayesian_2022} under increasingly realistic conditions.

In many information engines, feedback is applied conditionally, depending on the outcome of a measurement\cite{goerlich2025experimental}. Rather than acting on every observed microstate, the engine waits for a favorable fluctuation before it extracts or stores work. The performance of such engines therefore depends not only on the work extracted per successful action, but also on how frequently these favorable fluctuations can be detected. A natural measure of performance is the mean extracted power, which involves a trade-off between the work extracted in each successful cycle and the time required to complete that cycle. This trade-off becomes more intricate when the cost associated with acquiring and processing information is considered. Beyond the fundamental thermodynamic cost of information erasure~\cite{landauer1961irreversibility, berut2012experimental}, the energetic expenditures associated with sensing, processing, and feedback can influence the optimal operation of information engines. The resulting trade-off between detection time and measurement cost makes the measurement interval an important control parameter. Optimizing finite-time performance therefore requires understanding the temporal statistics of favorable fluctuations.

The first-passage time to a favorable fluctuation depends strongly on correlations between successive measurements. When the measurement interval exceeds the characteristic decorrelation time of the system, successive outcomes may be treated as approximately independent. In this limit, the number of attempts required to observe a favorable fluctuation follows a geometric distribution, yielding a mean waiting time
\begin{equation}
    \langle T_{\textrm{FP}}\rangle_{\textrm{ind}} = \frac{\tau_m}{p_1},
    \label{eq:mean_fp_ind}
\end{equation}
where $\tau_m$ is the measurement interval and $p_1$ is the stationary probability of a successful measurement~\cite{remi_now, hoek2026piston}. Measurements are performed at times $\tau_m,\:2\tau_m,\:\ldots$, with the waiting time obtained from the beginning of the observation process. In systems with persistent dynamics, however, consecutive measurements are correlated with previous outcomes~\cite{Garcia-Millan2025Optimal, admon_experimental_2018}. A sequence of unsuccessful measurements may therefore persist much longer than predicted by the independent-measurement approximation, even when the stationary success probability is unchanged. The waiting-time statistics depend not only on how often favorable fluctuations occur, but also on how they are organized in time. This raises a central question: Do prediction and optimal feedback require the same knowledge of temporal correlations?

Recent work has addressed related aspects of this problem. Costly measurements can produce nontrivial measurement schedules in mesoscopic information engines~\cite{Panizon2026Optimal}. Experiments in non-Markovian environments have also shown that hidden environmental memory can be accessed and converted into additional work~\cite{Muruga2026Hidden}. Here we ask a complementary question: how much of the temporal structure of a correlated active bath must a controller retain to choose a near-optimal feedback protocol?

Active matter provides a natural platform for studying temporal correlations in feedback-driven engines. Unlike passive particles governed by thermal fluctuations, active particles continuously consume energy to sustain their motion, driving the system out of equilibrium\cite{fodor2016far} and producing rich phenomena such as flocking~\cite{vicsek2012collective, Liebchen2017}, motility-induced phase separation~\cite{cates2015motility, burekovic2026bulk}, and giant number fluctuations~\cite{Deseigne2010, Omer_2023}. These nonequilibrium dynamics challenge conventional thermodynamic descriptions~\cite{Solon2015} and generate nontrivial spatial and temporal statistics. In particular, the persistence of self-propelled motion introduces memory that can extend over timescales comparable to, or longer than, the interval between measurements $\tau_m$. Under confinement, persistent particles tend to accumulate near boundaries, where they remain until their orientation changes sufficiently to allow them to escape~\cite{elgeti2013wall, fily2014dynamics}. This leads to strongly inhomogeneous spatial distributions, with boundary accumulation shaping both the spatial and temporal characteristics of fluctuations in the active gas \cite{hoek2026piston}. In active systems, the continuous energy consumption required to sustain particle motion is typically treated as a given, with housekeeping dissipation excluded from the effective work balance. Measurement-based feedback enables the extraction of useful work from this continuous energy input.

In a recent study, we introduced a macroscopic piston-like information engine containing a gas of self-propelled bristle-robots~\cite{hoek2026piston}. The engine periodically measures whether a region of width $\Delta x$ next to the piston is empty. If the region is empty, with probability $p_1(\Delta x)$, the piston advances by $\Delta x$ at no work cost. This compresses the gas and stores energy that can later be extracted during expansion. If the region is occupied, the piston remains stationary until a favorable fluctuation is detected. Our previous work showed that boundary accumulation changes the probability of these fluctuations and shifts the optimal operating protocol.

Here, we extend this picture to the temporal domain and ask how much temporal information is needed for effective control. Experimentally measured vacancy fluctuations reveal first-passage statistics that deviate strongly from independent-measurement predictions and are not fully captured by a first-order Markov approximation. This reduced description still selects protocols that achieve nearly the maximum experimentally attainable power. With a finite cost per measurement, the optimum shifts from frequent to sparse sensing, and the first-order Markov approximation remains nearly optimal when tested on independent segments of the data. Our results show that higher-order memory is important for predicting the full first-passage dynamics, while a simpler one-step description can be sufficient for near-optimal fixed-interval feedback control.

\section{Experimental System and Measurement Protocol}\label{sec:2-Experimental}

\begin{figure}[b]
    \centering
    \includegraphics[width=\linewidth]{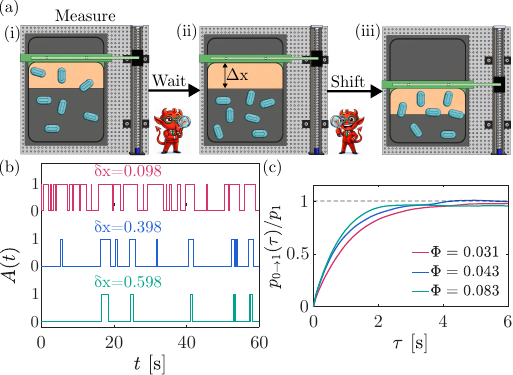}
    \caption{(a) Schematic of the operation protocol of the piston-like information engine containing self-propelled bristle-robots. (i) The engine monitors a region of width $\Delta x$ adjacent to the piston. (ii) Measurements are repeated until the monitored region is empty, allowing the piston to (iii) advance by $\Delta x$ without directly pushing the particles. (b) Representative binary signals $A(t)$ for three normalized compression widths, $\delta x=\Delta x/L_x$, at fixed particle density $\Phi = 0.053$. The signal takes the value $A=1$ when the monitored region is empty and $A=0$ otherwise. (c) Normalized conditional transition probability $p_{0\to1}(\tau)/p_1$ as a function of the time interval $\tau$ between measurements, for three packing fractions $\Phi$ at fixed compression width $\delta x = 0.398$. Here, $p_{0\to1}(\tau)$ is the probability of observing an empty region after an occupied observation, and $p_1$ is the stationary empty-region probability. The dashed horizontal line indicates the independent-measurement limit, $p_{0\to1}/p_1=1$.}
    \label{fig:Fig1}
\end{figure}

We consider the macroscopic piston-like information engine introduced in Ref.~\cite{hoek2026piston}, consisting of a two-dimensional gas of self-propelled bristle-robots (Hexbug nano$^{\text{\textregistered}}$) confined within an arena by rigid walls and a movable piston [Fig.~\ref{fig:Fig1}(a)]. The particles are driven by internal rotation motors, generating persistent chiral motion. The experimental apparatus and particle dynamics are described in detail in Ref.~\cite{hoek2026piston}, while the experimental parameters and trajectory-processing procedures relevant to the present analysis are summarized in Appendix A. The engine protocol is shown in Fig.~\ref{fig:Fig1}(a). Measurements are performed at regular intervals $\tau_m$, and a successful vacancy detection triggers a piston displacement $\Delta x$. The piston then expands under the pressure of the active gas and returns to its initial position. The two control parameters are therefore the compression distance $\Delta x$ and the measurement interval $\tau_m$.

In the present study, we evaluate feedback protocols retrospectively using experimentally recorded particle trajectories. The extracted first-passage statistics are combined with independently characterized piston expansion dynamics to calculate the expected cycle duration and power output for each protocol.

To characterize the measurement outcomes, we construct a binary occupancy signal $A(t,\Delta x)$ from the experimentally recorded particle trajectories, defined as
\begin{equation}
A(t,\Delta x)=
\begin{cases}
1 & \text{if the monitored region is empty},\\
0 & \text{otherwise}.
\end{cases}
\label{eq:vacancy_signal}
\end{equation}
The stationary probability of a successful measurement is then given by
\begin{equation}
p_1(\Delta x)=\langle A(t,\Delta x)\rangle_t,
\label{eq:p1}
\end{equation}
where $\langle\cdot\rangle_t$ denotes a time average. Representative occupancy signals for three normalized compression widths, $\delta x=\Delta x/L_x$, are shown in Fig.~\ref{fig:Fig1}(b), where $L_x$ is the initial length of the arena. As the monitored region ($\Delta x$) increases in size, empty-region events become less frequent, while occupied intervals generally become longer. Although $p_1$ determines the fraction of successful measurements, it does not describe how these events are distributed in time. Consequently, occupancy signals with the same stationary probability may exhibit substantially different waiting-time statistics.

To quantify correlations between successive measurements, we consider the conditional transition probability
\[
p_{0\to1}(\tau,\Delta x)=
P\!\left[A(t+\tau,\Delta x)=1\,\middle|\,A(t,\Delta x)=0\right].
\]
It gives the probability of finding the monitored region empty after a delay $\tau$, given that it was occupied initially. For independent measurements, $p_{0\to1}=p_1$. More generally, this transition probability is related to the normalized autocorrelation function through
\[
\frac{p_{0\to1}(\tau,\Delta x)}{p_1(\Delta x)}
=1-C_A(\tau,\Delta x).
\]
Figure~\ref{fig:Fig1}(c) shows that this ratio is strongly suppressed at short times, reflecting the persistence of occupied states. It approaches unity as temporal correlations decay.  These correlations have direct consequences for engine operation, because the piston advances only upon detecting an empty region; the sequence of measurement outcomes governs the waiting time for a successful feedback step. This waiting time contributes to the total cycle duration and therefore affects the extracted power. To characterize it, we turn to the first-passage statistics of the occupancy signal, which describe the time until the first successful measurement.

\section{Temporal Correlations Control First-Passage Statistics}

The finite-time operation of the engine is governed by the time required to detect an empty region and perform a successful feedback step. To characterize this waiting time, we consider the first-passage statistics of the occupancy signal $A(t,\Delta x)$, sampled at discrete intervals $\tau_m$. Starting from an occupied region, $A(0)=0$, we define the first-passage time $T_{\mathrm{FP}}$ as the time elapsed until the first subsequent successful measurement, $A=1$. The corresponding survival probability is defined as
\begin{equation}
S(t)=P(T_{\mathrm{FP}}>t\mid A(0)=0),
\label{eq:survival}
\end{equation}
representing the probability that no successful measurement has occurred by time $t$. Since measurements are performed at discrete intervals $\tau_m$, the first-passage time takes values $T_{\mathrm{FP}}=n\tau_m$, where $n=1,2,\ldots$. Its mean is related to the survival probability through
\begin{equation}
\langle T_{\mathrm{FP}}\rangle
=\tau_m\sum_{n=0}^{\infty}S(n\tau_m).
\label{eq:mean_first_passage}
\end{equation}
Both quantities are extracted directly from the experimental occupancy time series, without assuming independent measurement outcomes. First-passage starting points are separated sufficiently in time to ensure that the resulting samples are statistically independent.

We first consider the limit of statistically independent measurements, where each observation has a probability $p_1$ of detecting an empty region, regardless of previous outcomes. Since each measurement is independent, the survival probability after $n$ unsuccessful attempts is given by
\begin{equation}
S_{\mathrm{ind}}(n\tau_m)=(1-p_1)^n.
\label{eq:survival_ind}
\end{equation}
Substituting this expression into Eq.~\eqref{eq:mean_first_passage} recovers the independent mean waiting time introduced in Eq.~\eqref{eq:mean_fp_ind}: $\langle T_{\mathrm{FP}}\rangle_{\mathrm{ind}}=\tau_m/p_1$. This result provides a baseline for quantifying the effects of temporal correlations on the engine's waiting-time statistics.

To account for temporal correlations, we next consider a first-order Markov approximation, in which the probability of each measurement outcome depends only on the immediately preceding observation. Within this description, the probability of detecting an empty region following an unsuccessful measurement is given by the experimentally determined transition probability $p_{0\to1}(\tau_m)$. Since each unsuccessful measurement leaves the engine in the occupied state, the survival probability after $n$ consecutive attempts becomes
\begin{equation}
S_{\mathrm{M}}(n\tau_m)=
\left[1-p_{0\to1}(\tau_m)\right]^n.
\label{eq:survival_markov}
\end{equation}
The corresponding mean first-passage time is
\begin{equation}
\langle T_{\mathrm{FP}}\rangle_{\mathrm{M}}
=\frac{\tau_m}{p_{0\to1}(\tau_m)}.
\label{eq:mean_fp_markov}
\end{equation}
Unlike the independent approximation, this model accounts for the persistence of occupied states through their measured transition probability. However, it assumes that the probability of detecting an empty region is independent of how many consecutive unsuccessful measurements have preceded it. Deviations between the Markov prediction and experimental first-passage statistics therefore reveal temporal dependencies extending beyond the most recent observation.

\begin{figure}
    \centering
    \includegraphics[width=\linewidth]{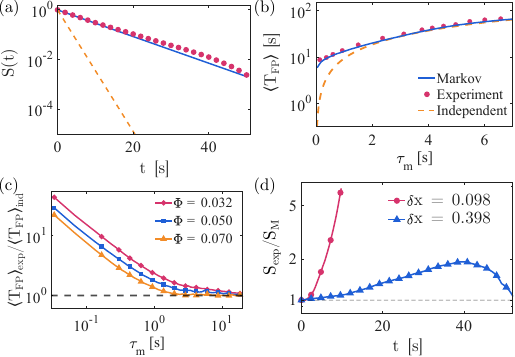}
    \caption{(a) Representative first-passage survival probability $S(t)$, giving the probability of observing no empty-region states up to time $t$, given that the region was initially occupied. Experimental results (pink circles) are compared with the independent-measurement approximation (orange dashed line) and the first-order Markov model (blue solid line) for $\Phi = 0.053$, $\delta x = 0.398$, and $\tau_m = 1/30$\thinspace{s}. (b) Mean first-passage time $\langle T_{FP}\rangle$ as a function of the measurement interval $\tau_m$, comparing experimental measurements with the same two statistical models as panel (a). (c) Ratio of experimental to independent mean first-passage times, $\langle T_{\mathrm{FP}}\rangle_{\mathrm{exp}}/\langle T_{\mathrm{FP}}\rangle_{\mathrm{ind}}$, for three packing fractions $\Phi$, at fixed compression width $\delta x = 0.398$. The dashed horizontal line denotes the independent limit, $\langle T_{\mathrm{FP}}\rangle_{\mathrm{exp}}/\langle T_{\mathrm{FP}}\rangle_{\mathrm{ind}}=1$. (d) Ratio of experimental to Markov survival probabilities, $S_{\mathrm{exp}}(t)/S_{\mathrm{M}}(t)$, for two normalized compression widths, at fixed packing fraction $\Phi =0.053$ and measurement interval $\tau_m = 0.2$\thinspace{s}. The dashed line denotes exact agreement with the Markov prediction.}
    \label{fig:2}
\end{figure}

Figure~\ref{fig:2}(a) compares the experimental first-passage survival probability with the independent and first-order Markov predictions. The independent approximation substantially underestimates the survival probability. The first-order Markov approximation captures the overall decay but deviates at longer times. The same pattern appears in the mean first-passage time [Fig.~\ref{fig:2}(b)]: the independent approximation predicts much shorter waiting times at small measurement intervals, while the first-order Markov prediction remains considerably closer to the experimental values. As $\tau_m$ increases, both approximations converge toward the experimental results, consistent with the decay of temporal correlations.

The magnitude of these correlation effects is quantified in Fig.~\ref{fig:2}(c) through the ratio of empirical to independent mean first-passage times for several packing fractions $\Phi$. At short measurement intervals, this ratio exceeds unity by more than an order of magnitude, demonstrating the substantial waiting-time penalty associated with correlated sampling. Figure~\ref{fig:2}(d) further reveals the limitations of the Markov approximation: the increasing ratio $S_{\mathrm{exp}}(t)/S_{\mathrm{M}}(t)$ indicates that long occupied intervals persist more frequently than predicted, particularly at smaller compression distances. This behavior is consistent with persistent boundary accumulation, where particles may remain trapped against the piston until their orientations allow escape\cite{fily2014dynamics}. Since the binary occupancy signal does not resolve particle orientations or configurations, these hidden degrees of freedom can generate memory beyond the most recent measurement, providing a physical interpretation of the observed higher-order temporal dependencies.

\section{Temporal Correlations Shift the Optimal Engine Protocol}

\begin{figure}[t]
    \centering
    \includegraphics[width=\linewidth]{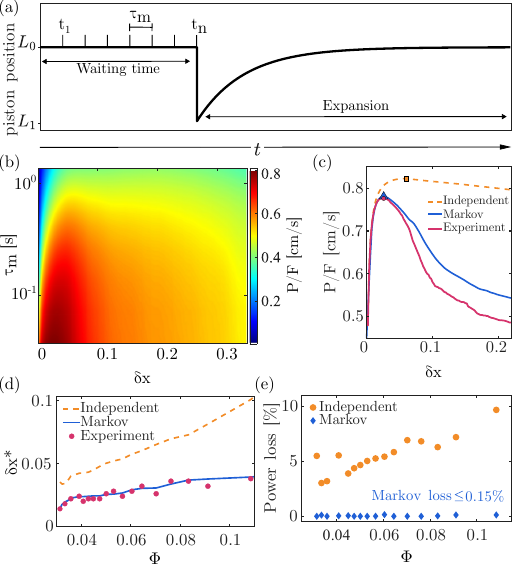}
    \caption{(a) Schematic of an engine cycle, showing repeated measurements at intervals $\tau_m$ until the monitored region is found empty. A successful measurement triggers an instantaneous piston displacement $\Delta x$, followed by expansion toward its initial position. (b) Experimental power landscape $P/F$ as a function of normalized compression width and measurement interval $\tau_m$, at fixed packing fraction $\Phi = 0.053$. Here, $F$ is the force opposing piston expansion. (c) Power as a function of $\delta x$ at fixed $\Phi=0.053$ and $\tau_m = 1/30$\thinspace{s}, comparing experimental first-passage statistics (pink) with the first-order Markov (blue) and independent-measurement (orange dashed) predictions. Symbols indicate the corresponding optimal compression widths. (d) Optimal normalized compression width $\delta x^*$ as a function of packing fraction $\Phi$, obtained by maximizing the power over compression width and measurement interval for each model. (e) Predicted power loss resulting from operating at the protocols selected by the independent and Markov models, relative to the experimentally optimized protocol. Neglecting temporal correlations produces power losses of several percent, whereas the Markov-selected protocols retain nearly all the experimentally achievable power, with losses not exceeding $0.15\%$.}
    \label{fig:Fig3}
\end{figure}

We now examine how temporal correlations affect the engine power and optimal protocol. Each cycle consists of repeated measurements until an empty region is detected [Fig.~\ref{fig:Fig3}(a)]. The piston then advances by $\Delta x$ and expands back to its initial position. We assume that the active gas has relaxed to its stationary state before each new cycle. The first measurement therefore succeeds with probability $p_1(\Delta x)$. If it fails, the cycle includes an additional mean waiting time $\langle T_{\mathrm{FP}}\rangle$. The mean cycle duration is
\begin{equation}
\langle T_{\mathrm{cycle}}\rangle
=\tau_m+[1-p_1(\Delta x)]\langle T_{\mathrm{FP}}\rangle
+t_{\mathrm{exp}}(\Delta x).
\label{eq:cycle_duration}
\end{equation}
 The expansion time $t_{\mathrm{exp}}(\Delta x)$ is obtained from independent measurements of the piston dynamics reported in Ref.~\cite{hoek2026piston}. The piston expansion time is much longer than the measured occupancy correlation time, so the active gas effectively relaxes before the next cycle begins. The mean piston displacement follows $\langle\Delta X(t)\rangle=\Delta X_{\max}(1-e^{-t/\tau_{\mathrm{exp}}^{(0)}})$, with $\Delta X_{\max}=42.10\pm0.04$\thinspace{cm} and $\tau_{\mathrm{exp}}^{(0)}=48.7\pm0.2$\thinspace{s}. The corresponding expansion time is therefore
\begin{equation}
t_{\mathrm{exp}}(\Delta x)=
-\tau_{\mathrm{exp}}^{(0)}
\ln\left(1-\frac{\Delta x}{\Delta X_{\max}}\right).
\label{eq:expansion_time}
\end{equation}
Each completed cycle extracts mechanical work $W=F\Delta x$ against an opposing force $F$, yielding a mean power
\begin{equation}
P(\Delta x,\tau_m)
=\frac{F\Delta x}{\langle T_{\mathrm{cycle}}\rangle}.
\label{eq:engine_power}
\end{equation}
The engine's power reflects a competition between the work gained from larger compressions and the associated increase in cycle duration. Temporal correlations modify this balance through their influence on the waiting time for a successful measurement, potentially shifting the optimal operating conditions.

Figure~\ref{fig:Fig3}(b) shows the experimental power landscape as a function of the normalized compression distance $\delta x=\Delta x/L_x$ and measurement interval $\tau_m$, at $\Phi=0.053$. Increasing the compression distance increases the work extracted per cycle, but reduces the probability of detecting an empty region, leading to longer waiting times. This competition produces a power maximum at an intermediate compression distance. Temporal correlations shift its location. Frequent measurements probe strongly correlated configurations, so reducing $\tau_m$ does not proportionally reduce the waiting time for a vacancy. Nevertheless, for nested measurement intervals, finer sampling can only detect a vacancy earlier or at the same time. Consequently, in the absence of measurement costs, power is necessarily maximized at the shortest available interval, $\tau_m=1/30$\thinspace{s}. Temporal correlations therefore influence the power output and optimal compression distance, while finite measurement costs introduce a nontrivial optimization of the measurement interval.

To isolate the influence of temporal correlations on the optimal compression distance, we compare the empirical power with predictions based on independent and first-order Markov measurements. Figure~\ref{fig:Fig3}(c) shows that the independent approximation overestimates the power and favors larger compression distances than those observed experimentally. By neglecting correlations, this approximation underestimates the waiting time for a successful measurement. Larger compressions therefore appear more favorable than they are, despite their lower vacancy probabilities. Incorporating the measured transition probability $p_{0\to1}(\tau_m)$ restores much of this waiting-time penalty, shifting the predicted optimum toward smaller compression distances. As shown in Fig.~\ref{fig:Fig3}(d), this distinction persists across the investigated packing fractions: the Markov approximation closely tracks the experimental optimal compression distance $\delta x^*$, while the independent model systematically favors larger values. Temporal correlations therefore affect the optimal operating conditions beyond the spatial inhomogeneity already captured by the stationary vacancy probability $p_1(\Delta x)$.

To assess the effect of these differences, we evaluate the experimental power at the operating conditions selected by each approximation. We compare this with the maximum obtained from the empirical first-passage statistics. Figure~\ref{fig:Fig3}(e) shows that neglecting temporal correlations leads to power losses of several percent. The first-order Markov approximation gives much smaller losses, about $0.15\%$ or less across the investigated packing fractions. Its prediction of the mean cycle duration remains accurate near the optimal compression distance. Larger errors appear mainly away from the high-power region, as shown in Appendix~\ref{App:B}, Fig.~\ref{fig:markov_cycle_accuracy}. Thus, the full first-passage dynamics are not needed to achieve near-optimal feedback control. A renewal description incorporating the measured distribution of occupied-state residence times provides a more detailed account of first-passage statistics, but does not meaningfully improve the selected protocol or its performance (Appendix~\ref{App:higher_order}).

\section{Measurement Costs Favor Sparse Sensing}

\begin{figure}[b]
    \centering
    \includegraphics[width=\linewidth]{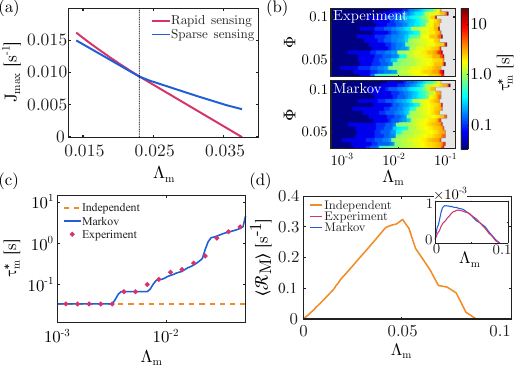}
    \caption{(a) Maximum normalized net-power objective $J_{\max}$ as a function of dimensionless measurement cost $\Lambda_m=\epsilon_m/(FL_x)$, for rapid ($\tau_m\leq0.5\,\mathrm{s}$) and sparse ($\tau_m>0.5\,\mathrm{s}$) sensing protocols at fixed packing fraction $\Phi = 0.053$. The vertical dotted line marks the cost at which sparse and rapid sensing become equal. The objective accounts for both extracted work and the cost of repeated measurements. (b) Optimal measurement interval $\tau_m^*$ as a function of packing fraction $\Phi$ and measurement cost $\Lambda_m$, obtained from experimental first-passage statistics (top) and the first-order Markov approximation (bottom). The color scale indicates $\tau_m^*$, while gray regions denote conditions under which no protocol yields positive net power. (c) Optimal measurement interval as a function of $\Lambda_m$ at fixed packing fraction $\Phi = 0.053$, comparing experiment (pink circles), Markov (blue solid line), and independent measurements (orange dashed line). While the independent model continues to favor the shortest available interval, $\tau_m=1/30\,\mathrm{s}$, the experimental and Markov optima shift toward increasingly sparse sensing as measurement costs increase. (d) Mean held-out regret $\langle\mathcal{R}_M\rangle$, defined as the difference between the experimentally optimal net-power objective and the experimentally evaluated performance of a model-selected protocol. Protocols are optimized using one temporal block and evaluated on another, with results averaged over both block assignments and densities. The inset magnifies the experimental and Markov regrets, which remain substantially smaller than those of the independent model.}
    \label{fig:Fig4}
\end{figure}

Thus far, we have treated measurements as cost-free, finding that the shortest available measurement interval maximizes the extracted power. In practice, however, sensing and information processing may incur energetic costs, introducing an additional trade-off between the frequency of measurements and the work extracted. To investigate this trade-off, we assign an effective cost $\epsilon_m$ to each measurement. This phenomenological parameter represents the energetic expenditure associated with sensing and processing and is not measured directly in the present experiment.
Defining the dimensionless measurement cost $\Lambda_m=\varepsilon_m/(FL_x)$, we introduce the cost-adjusted net-power measure
\begin{equation}
J(\delta x,\tau_m;\Lambda_m)=
\frac{\delta x-\Lambda_m\langle N\rangle}
{\langle T_{\mathrm{cycle}}\rangle},
\label{eq:net_power}
\end{equation}
where $\delta x=\Delta x/L_x$ and $\langle N\rangle$ is the mean number of measurements performed per cycle. Since each cycle begins with one measurement, followed by additional measurements only if the initial outcome is unsuccessful,
\begin{equation}
\langle N\rangle=
1+(1-p_1)\frac{\langle T_{\mathrm{FP}}\rangle}{\tau_m}.
\label{eq:mean_measurements}
\end{equation}
The first-passage statistics determine both the cycle duration and the accumulated measurement cost. Increasing the measurement interval can reduce the number of measurements required per cycle, but may also prolong the waiting time for a successful observation. The optimal sensing strategy follows from the balance between measurement cost and the temporal dynamics of the active gas.

Figure~\ref{fig:Fig4}(a) illustrates the competition between detection time and sensing expenditure by comparing the maximum net-power measure within rapid ($\tau_m\leq0.5$ s) and sparse ($\tau_m>0.5$ s) sensing regimes, with the compression distance and measurement interval optimized within each regime. At low measurement costs, frequent sampling minimizes detection delays, whereas increasing $\Lambda_m$ favors protocols that reduce the accumulated sensing cost. The crossing of the two branches identifies the cost at which the preferred regime changes. It does not imply a discontinuity in the optimal measurement interval.

The optimal measurement interval also depends on packing fraction, as shown by the experimental and Markov landscapes in Fig.~\ref{fig:Fig4}(b). This density dependence reflects variations in the temporal statistics of the active gas, while the agreement between the two landscapes indicates that the Markov approximation captures the main features relevant to sensing optimization. At fixed packing fraction [Fig.~\ref{fig:Fig4}(c)], the empirical and Markov optima shift toward longer intervals as measurement costs increase, whereas the independent approximation necessarily favors the shortest available interval whenever operation is profitable. Indeed, for independent measurements, $\langle N\rangle=1/p_1$, yielding
\begin{equation}
J_{\mathrm{ind}}=
\frac{\delta x-\Lambda_m/p_1}
{\tau_m/p_1+t_{\mathrm{exp}}(\Delta x)}.
\end{equation}
For positive net power, this expression decreases monotonically with $\tau_m$. The independent approximation cannot favor sparse sensing because it neglects temporal correlations, not because of the particular parameter values used here. At sufficiently high measurement costs, all protocols can become unprofitable, making no operation the preferred choice.

To test the robustness of these control strategies, we use separate temporal segments of the recorded trajectories. For each packing fraction, the time series is divided into two non-overlapping segments. Protocols are optimized using statistics from one segment and evaluated using empirical statistics from the other. We quantify performance by the regret $\mathcal{R}_M$, defined as the difference between the maximum experimental net power and the net power obtained by the selected protocol on the evaluation segment. Figure~\ref{fig:Fig4}(d) shows the mean regret, averaged over packing fractions and both choices of optimization and evaluation segments.

Neglecting temporal correlations leads to substantial losses, especially at higher measurement costs, where frequent sensing can become unfavorable. The first-order Markov approximation yields regret comparable to that of the empirically optimized strategy, despite its incomplete description of first-passage statistics. Thus, accurate reconstruction of the full temporal dynamics is not needed for effective feedback control. Dependence between successive measurements is sufficient to identify nearly optimal fixed-interval protocols, even when sensing has a finite cost.

The larger regret of the independent approximation has a simple origin. Protocols chosen under the assumption of uncorrelated measurements can incur substantial sensing costs when evaluated against experimental first-passage statistics. In some cases, this leads to negative net power. Because regret is measured relative to the best nonnegative performance, these losses can exceed the maximum positive net power.

\section{Discussion and Outlook}
Our results show that temporal correlations play a central role in the finite-time operation of active information engines. While the stationary probability of a favorable fluctuation determines how frequently the monitored region is empty, it does not capture the temporal organization of these events or the waiting time required to detect them. In our previous work, we established how boundary accumulation and the resulting spatial inhomogeneity reshape the probability of successful measurements and the optimal compression distance. Here, we show that temporal correlations introduce an additional constraint on engine performance, substantially modifying first-passage times and shifting the optimal operating conditions. Together, these results show how spatial and temporal statistics jointly determine the use of nonequilibrium fluctuations for work extraction.

The distinction between statistical prediction and optimal control follows from the dependence of mean power on the waiting-time statistics. Predicting the full first-passage distribution requires a detailed description of temporal correlations. Power optimization depends mainly on the mean cycle duration and on how it varies with the control parameters. The first-order Markov approximation captures this dependence sufficiently well near optimal operation, allowing it to identify nearly optimal protocols. This connects to recent work on optimal control of active particles, where feedback exploits information about self-propulsion to improve performance~\cite{Garcia-Millan2025Optimal}. Here, the controller has access only to a binary occupancy signal from a many-particle active bath. Our results show that even this reduced description can be sufficient for near-optimal control. More generally, the information needed to control an active system can be simpler than the information needed to describe its dynamics.

Finite measurement costs reveal another consequence of temporal correlations: sparse sensing can become optimal. For independent measurements, the expected number of observations per successful cycle is independent of the measurement interval, so increasing this interval only prolongs the cycle without reducing its expected sensing cost. Temporal correlations fundamentally alter this trade-off, as closely spaced measurements repeatedly probe persistent configurations. Increasing the interval can therefore reduce sensing expenditure sufficiently to compensate for longer detection times. The first-order Markov approximation captures this cost-dependent shift. Even a minimal representation of temporal memory can therefore guide sensing optimization.

The present optimization is restricted to protocols with fixed compression distance and fixed measurement interval within each cycle. The observed higher-order temporal memory may become useful for more adaptive strategies, in which the measurement interval depends on the duration of the preceding occupied sequence. Such protocols could reduce redundant measurements during persistent occupation. Testing whether they outperform the fixed-interval strategies considered here is a natural extension.

\begin{acknowledgments}
Y.R. and L.H. acknowledge support from the European Research Council (ERC) under the European Union’s Horizon 2020 research and innovation program (Grant Agreement No. 101002392).
\end{acknowledgments}

\section*{AUTHOR DECLARATIONS}
\subsection*{Conflict of Interest}
The authors have no conflicts to disclose.
\subsection*{Author Contributions}

\textbf{Laura Hoek:}
Conceptualization (lead);
Methodology (lead);
Investigation (lead);
Data curation (lead);
Formal analysis (lead);
Visualization (lead);
Writing - original draft (lead);
Writing - review \& editing (supporting).
\textbf{Yael Roichman:}
Conceptualization (supporting);
Supervision (lead);
Funding acquisition (lead);
Writing - review \& editing (lead).

\section*{DATA AVAILABILITY}
The data that support the findings of this study
are available from the corresponding author
upon reasonable request.

\appendix

\section{Experimental Details}

\subsection{Experimental System and Particle Tracking}

The experimental system consists of self-propelled bristle-robots confined within a two-dimensional rectangular arena of dimensions$L_x\times L_y$ with $10\leq L_x\leq 40$\thinspace{cm} and $L_y=29.2$\thinspace{cm}, bounded by rigid walls and a movable piston. The particles are propelled by internal-rotation motors, generating persistent chiral motion. The packing fraction is defined as $\Phi = N_b A_b/(L_x L_y)$, where $N_b=6$ is the particle number, and $A_b = 4\times 1.2$\thinspace{cm$^2$} is the projected area of an individual particle. Different packing fractions were obtained by varying the arena dimensions. Further details of the experimental apparatus and its mechanical characterization are provided in Ref.\cite{hoek2026piston}.

Particle trajectories were extracted from overhead video recordings using colored markers attached to the bristle-robots. The recorded positions were analyzed at a frame rate of 30 fps, providing a temporal resolution of $1/30$\thinspace{s}. The resulting trajectories were used to characterize the stationary spatial distributions and temporal fluctuations of the active gas under confinement.

\subsection{Binary occupancy signal and measurement protocol}

For each recorded trajectory, we constructed the binary occupancy signal $A(t,\, \Delta x)$ by determining whether a rectangular region of width $\Delta x$ adjacent to the piston was empty or occupied at each frame. The analysis was repeated for different normalized compression distances $\delta x = \Delta x/L_x$, allowing the stationary vacancy probability $p_1(\Delta x)$ and temporal statistics to be extracted from the same experimental recordings.

To investigate different measurement protocols, the binary signals were sampled at intervals of $\tau_m$, where $\tau_m$ is an integer multiple of the frame duration. This allowed the first-passage statistics to be evaluated across measurement intervals without changing the underlying particle dynamics. The resulting statistics were subsequently combined with the independently characterized piston expansion dynamics to calculate the expected cycle duration and power output for each operating protocol. Details of the first-passage analysis and its statistical limitations are provided in Appendix B.

\section{First-Passage Statistics and Statistical Models}\label{App:B}

\subsection{Extraction of experimental first-passage statistics}

First-passage statistics were obtained from the experimentally measured binary occupancy signal $A(t,\, \Delta x)$, defined in Sec.~\ref{sec:2-Experimental}. For each compression distance $\Delta x$ and measurement interval $\tau_m$, the signal was sampled at discrete intervals of $\tau_m$. Starting from an occupied observation, $A(0) = 0$, the first-passage time $T_{\mathrm{FP}}$ was defined as the elapsed time until the first subsequent observation of an empty region, $A(n\tau_m) = 1$.

The survival probability $S(n\tau_m)$ represents the probability that no successful measurement occurs during the first n observations following an initially occupied state. For each initially occupied observation, the first-passage time was determined from the subsequent sampled occupancy sequence, and the empirical survival probability was obtained from the resulting waiting-time statistics.

The stationary vacancy probability $p_1(\Delta x)$ and conditional transition probability $p_{0\to1}(\tau_m,\,\Delta x)$ were extracted from the same occupancy signals. These quantities serve as inputs to the independent and first-order Markov approximations, respectively, allowing their predictions to be compared with the experimental first-passage statistics without introducing additional fitted parameters.

\subsection{First-order Markov approximation}
To incorporate correlations between successive observations, we consider a first-order Markov approximation. The binary occupancy signal is represented by two states, occupied (0) and empty (1), with transitions described by the matrix

\begin{equation}
\mathbf{M}(\tau_m)=
\begin{pmatrix}
1-a & a\\
b & 1-b
\end{pmatrix},
\end{equation}

where $a = p_{0\to1}(\tau_m)$ and $b = p_{1\to0}(\tau_m)$ are the conditional transition probabilities. The rows and columns correspond to the occupied and empty states, respectively.

The Markov assumption states that the probability of the next measurement outcome depends only on the current state, independently of the preceding measurement history. Since the first-passage process begins in state 0 and terminates upon reaching state 1, survival requires the system to remain in state 0 at every sampled observation. Consequently,

\begin{equation}
    S_{\mathrm M}(n\tau_m)=(1-a)^n,
\end{equation}

and the corresponding first-passage distribution is

\begin{equation}
    P_{\mathrm M}(T_{\mathrm{FP}}=n\tau_m)
    =a(1-a)^{n-1}.
\end{equation}

The mean first-passage time is therefore

\begin{equation}
\langle T_{\mathrm{FP}}\rangle_{\mathrm M}
=\tau_m\sum_{n=0}^{\infty}(1-a)^n
=\frac{\tau_m}{a}.
\end{equation}

The measured transition probability is directly related to the normalized autocorrelation function of the binary occupancy signal through

\begin{equation}
p_{0\to1}(\tau_m)
=p_1[1-C_A(\tau_m)].
\end{equation}

Substituting this relation yields

\begin{equation}
\langle T_{\mathrm{FP}}\rangle_{\mathrm M}
=\frac{\tau_m}{p_1[1-C_A(\tau_m)]}.
\end{equation}

This expression explicitly connects temporal correlations to the predicted waiting time. In the independent limit, $C_A(\tau_m) = 0$, the Markov prediction reduces to the independent result.

The distinction between these approximations lies in the information retained about previous measurements. The independent model uses only the stationary vacancy probability, whereas the Markov approximation additionally incorporates the conditional probability of a transition from one observation to the next. Although the latter reproduces the measured single-step transition probabilities by construction, it does not account for dependencies on earlier outcomes. Such higher-order temporal dependencies can therefore produce deviations between the experimental and Markov first-passage distributions.

\subsection{Higher-order temporal memory}\label{App:higher_order}
To examine the higher-order temporal dependencies revealed by the first-passage statistics, we also considered a renewal description based on the distribution of consecutive occupied observations. Unlike the first-order Markov approximation, which assigns a constant transition probability $p_{0\to1}(\tau_m)$ after each unsuccessful measurement, this description retains the distribution of residence times in the occupied state.

For a fixed compression distance $\Delta x$ and measurement interval $\tau_m$, we identified consecutive runs of occupied observations, $A=0$, in the sampled binary occupancy signal. Let $q(k)$ denote the probability that an occupied run contains $k$ consecutive observations. In a stationary time series, an observation selected at random from the occupied state is more likely to belong to a long run than to a short one. The residual duration of the occupied run therefore determines the corresponding first-passage survival probability,
\begin{equation}
    S_{\mathrm{ren}}(n\tau_m) = \frac{1}{\langle k\rangle} \sum_{k=n+1}^{\infty}(k-n)q(k),
\end{equation}

where $\langle k\rangle=\sum_{k=1}^{\infty}kq(k)$ is the mean occupied-run length. The associated mean first-passage time is

\begin{equation}
   \langle T_{\mathrm{FP}}\rangle_{\mathrm{ren}} = \frac{\tau_m}{2}
\frac{\langle k^2\rangle+\langle k\rangle}
{\langle k\rangle}.
\end{equation}

For geometrically distributed occupied-run lengths, the renewal description reduces to the first-order Markov result, $\langle T_{\mathrm{FP}}\rangle_{\mathrm{M}}=\tau_m/p_{0\to1}(\tau_m)$. Deviations from a geometric residence-time distribution therefore provide a complementary characterization of the memory neglected by the Markov approximation.

Using the experimentally measured occupied-run distribution, the renewal description reconstructs the first-passage statistics while retaining the higher-order temporal information neglected by the first-order Markov approximation.

The renewal description also provides a natural perspective on possible extensions of the present control strategy. In particular, a controller that adjusts its measurement schedule according to the duration of an occupied sequence could potentially exploit information unavailable to the first-order Markov model. Whether such history-dependent protocols provide an operational advantage remains to be investigated.

\subsection{Accuracy of the Markov Approximation Near Optimal Operation}
To examine why the first-order Markov approximation identifies nearly optimal operating conditions despite discrepancies in first-passage statistics, we compare its predicted mean cycle duration with the mean obtained from empirical waiting times. Since both descriptions assign the same work to each successful cycle, their normalized powers satisfy

\begin{equation}
\frac{\langle T_{\mathrm{cycle}}\rangle_{\mathrm{exp}}}
{\langle T_{\mathrm{cycle}}\rangle_{\mathrm{M}}}
=
\frac{P_{\mathrm{M}}}{P_{\mathrm{exp}}}.
\end{equation}

\begin{figure}
    \centering
    \includegraphics[width=1\linewidth]{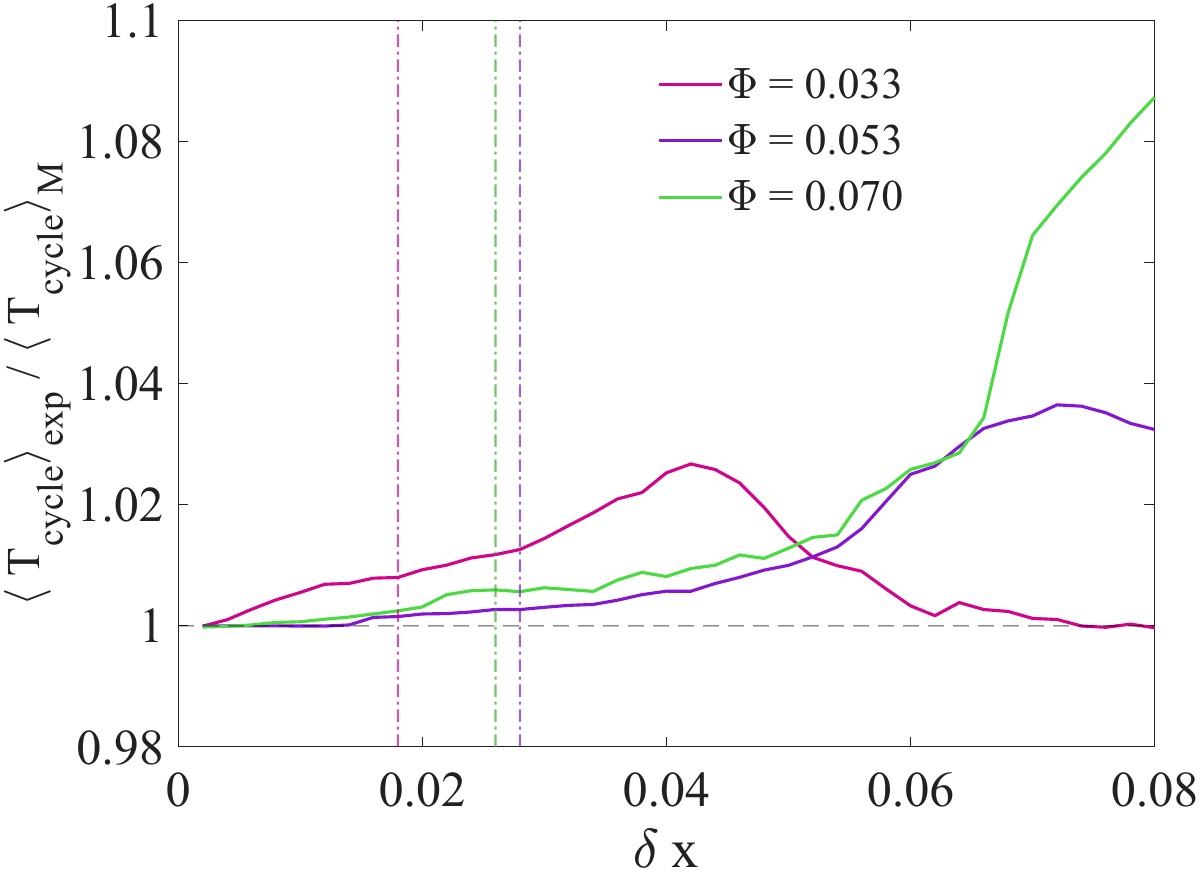}
    \caption{Ratio of empirical to Markov-predicted mean cycle durations as a function of normalized compression distance $\delta x$, at $\tau_m=1/30$ s, for three representative packing fractions. Vertical dotted lines indicate the corresponding empirically optimal compression distances. The two predictions agree closely near the optimal operating regime, while larger discrepancies emerge at greater compression distances. The dashed horizontal line indicates perfect agreement.}
    \label{fig:markov_cycle_accuracy}
\end{figure}

Figure~\ref{fig:markov_cycle_accuracy} shows this ratio as a function of normalized compression distance at the shortest measurement interval, $\tau_m=1/30$ s. The ratio remains close to unity at small compression distances, including the vicinity of the optimal operating points, but generally increases at larger compressions. Thus, the Markov approximation is most accurate in the operating regime relevant to power optimization, while larger discrepancies occur primarily away from the optimum. This explains, in part, why the approximation can identify nearly optimal protocols without reproducing the full first-passage statistics.

\subsection{Finite-record effects and statistical reliability}

The finite duration of experimental recordings limits the statistical accuracy of estimates of long first-passage times, particularly for large compression distances, where empty-region events are rare. For each measurement interval $\tau_m$, first-passage statistics were extracted from the corresponding subsampled binary occupancy signal. Waiting-time sequences that extended beyond the available recording could not be assigned a completed first-passage time. The resulting survival distributions were therefore evaluated only over observation times with sufficient statistical support, avoiding poorly sampled tails and unstable ratios between empirical and model survival probabilities.

The same considerations apply to power optimization, where poorly sampled first-passage times can lead to unreliable estimates of the mean cycle duration. We therefore restricted the optimization to parameter combinations with adequately sampled waiting-time statistics. All statistical models were evaluated over the same supported parameter range, ensuring that differences in their optimal protocols reflect their treatment of temporal correlations rather than differences in data availability.

\section{Cross-Validation and Regret Analysis}

\subsection{Cross-validation procedure}

To assess the robustness of the optimal feedback protocols and avoid evaluating their performance on the same data used for optimization, we performed a cross-validation analysis using nonoverlapping temporal segments of the experimental recordings. For each packing fraction, the recorded binary occupancy signal was divided into two consecutive blocks. The stationary vacancy probability and first-passage statistics were estimated separately from each block and used to construct the corresponding experimental, independent, and first-order Markov net-power landscapes.

For each statistical description, the optimal compression distance and measurement interval were determined by maximizing the net-power objective using the training block. The selected protocol was then evaluated using empirical first-passage statistics from the held-out block. This procedure was repeated with the roles of the two blocks reversed. All models were optimized over a common set of statistically supported protocols, including the option of no operation when the predicted net power was nonpositive.

For each training–test split, the admissible protocol set was restricted to compression distances and measurement intervals for which the held-out empirical net-power objective could be evaluated reliably. The same set was used for all three statistical descriptions, so that differences in regret arise from model-based protocol selection rather than unequal access to operating conditions.

\subsection{Regret and statistical evaluation}

We quantify the performance of each control strategy through its regret, defined as the difference between the maximum net-power objective obtained from held-out empirical statistics and the corresponding objective of the model-selected protocol. For a model M, this is expressed as
\begin{equation}
\begin{aligned}
\mathcal{R}_M ={}&
\max\left[0,
\max_{\delta x,\tau_m}
J_{\mathrm{exp}}^{\mathrm{test}}
(\delta x,\tau_m;\Lambda_m)\right]
\\
&-
J_{\mathrm{exp}}^{\mathrm{test}}
\left(\delta x_M^*,\tau_{m,M}^*;\Lambda_m\right),
\end{aligned}
\label{eq:regret}
\end{equation}

where the superscript $*$ denotes the protocol selected by maximizing the model's net-power objective on the training block, and the superscript test denotes evaluation using the held-out experimental data. If the model selects no operation, its evaluated objective is set to zero.

This definition measures the loss in empirically evaluated net power associated with a model-selected protocol, rather than the error in its predicted objective value. A model may therefore achieve low regret despite inaccurate power predictions if it identifies a protocol close to the empirical optimum.

The regret was evaluated for the experimental, independent, and first-order Markov descriptions across the investigated measurement costs and packing fractions. For each cost, the mean regret was obtained by averaging over the two assignments of training and evaluation blocks and over packing fractions satisfying the statistical reliability criteria. These averages are presented in Fig.~\ref{fig:Fig4}(d).

The empirically optimized strategy provides a benchmark for finite-sampling effects and differences between temporal segments. Its held-out regret need not vanish, since the optimal protocols identified from the training and test segments may differ. Comparing the Markov regret with this benchmark therefore reveals whether the approximation introduces additional performance losses beyond those associated with finite experimental statistics.

\bibliography{correlated_engine.bib}
\end{document}